\documentclass[12pt]{article}
\usepackage[english]{babel}
\usepackage{amsmath}
\usepackage{cite}
\begin{document}

\title{\bf{}BRST Lagrangian construction for spin-$\frac{3}{2}$ field in Einstein space}

\author{\sc I.L. Buchbinder${}^{a}$\thanks{joseph@tspu.edu.ru},
V.A. Krykhtin${}^{ab}$\thanks{krykhtin@tspu.edu.ru},
\\[0.5cm]
\it ${}^a$Department of Theoretical Physics,\\
\it Tomsk State Pedagogical University, 
\it Tomsk 634061, Russia\\[0.3cm]
\it ${}^b$Laboratory of Mathematical Physics,\\
\it Tomsk Polytechnic University,
\it Tomsk 634034, Russia}
\date{}

\date{}

\maketitle

\begin{abstract}
We explore a hidden possibility of BRST approach to higher spin
field theory to obtain a consistent Lagrangian for massive
spin-$\frac{3}{2}$ field in Einstein space. Also, we prove that in
the space under consideration the propagation of spin-$\frac{3}{2}$
field is hyperbolic and causal.
\end{abstract}


In this note we discuss the features of Lagrangian formulation for
spin-$\frac{3}{2}$ field on a curved spacetime in framework of BRST
approach \cite{brst-1,brst-2,0603212,0703049,09021471,09120611}. It
is well known that the Lagrangian formulation of the higher spin
fields in arbitrary external background can be contradictory. The
problem of consistent propagation of fields in different backgrounds
and their Lagrangian description is one of the problems of higher
spin field theory. Corresponding aspects of spin-$\frac{3}{2}$ field
was studied in enormous number of papers (see e.g.
\cite{x-files,Velo,Waldron} and the references therein). However,
practically all consistent formulations for spin-$\frac{3}{2}$ were
given or in flat or in AdS spaces\footnote{We do not discuss here
the supergravity  where consistency for massless spin-$\frac{3}{2}$
field is conditioned by supersymmetrical coupling to massless
spin-$2$ field (see e.g. \cite{BK}.}

BRST approach to higher spin field theories is a universal method
for derivation of the Lagrangians for such fields beginning with
on-shell relations, which define the higher spin fields (e.g. the
relations defining irreducible representations of the Poincare or
AdS groups). Following the general BRST-BFV construction we begin
with a closed constraint algebra for the theory and built the
Lagrangians. This approach yields consistent formulation for
massless and massive, bosonic and fermionic arbitrary spin-$s$
fields in constant curvature space, however for consistency it
demands the same space even for spin $s=1,2$ fields where the
Lagrangian formulations exist in an arbitrary Riemann space and an
Einstein space respectively. Such a puzzle was resolved in our paper
\cite{09120611} exploring some hidden possibility of the BRST
approach. In this paper we demonstrate that the same hidden
possibility exists for spin-$\frac{3}{2}$ field as well and allows
us to get the consistent spin-$\frac{3}{2}$ Lagrangian formulation
in arbitrary Einstein space.

We begin with a brief discussion of the main idea  proposed in
\cite{09120611}. The Lagrangian construction in the BRST approach
\cite{brst-1,brst-2,0603212,0703049,09021471}  was carried out for
arbitrary spin fields. The basic notions of this approach are the
Fock space vector $|\Psi_s\rangle$, corresponding to spin $s$ and
the nilpotent BRST charge $Q$. The equations of motion and gauge
transformations are written in the form $Q|\Psi_s\rangle=0$ and
$\delta|\Psi_s\rangle=Q|\Lambda_s\rangle$ respectively, with the
BRST operator $Q$ being the same for all spins. Nilpotency of the
BRST operator provides us the gauge transformations and fields
$|\Psi_s\rangle$ and $|\Psi_s\rangle+Q|\Lambda_s\rangle$ are both
physical. Since we consider all spins simultaneously, then from
$Q^2|\Lambda_s\rangle=0$ follows $Q^2=0$. But if we want to
construct Lagrangian for the field with a given value $s$ of spin,
then it is sufficient to require a weaker condition that the BRST
operator for given spin $Q_s$ is not nilpotent in operator sense but
will be nilpotent only on the specific Fock vector parameter
$|\Lambda_s\rangle$ corresponding to a given spin $s$,
$Q_s^2|\Lambda_s\rangle=0$ only and $Q_s^2\neq0$ on states of
general form. Just this point allows us to construct Lagrangian for
spin-$\frac{3}{2}$ field in Einstein space\footnote{An Einstein
space is defined by the relation
 $R_{\mu\nu}=const\cdot g_{\mu\nu}$ with Weyl tensor be arbitrary
(see e.g. \cite{1stein99fold}).}.



We begin with pointing out that there exist the consistent equations
of motion for spin-$\frac{3}{2}$ field in a space-time different
from AdS. It is well known that spin-$\frac{3}{2}$ field $\psi_\mu$
(the Dirac index is suppressed) will describe the irreducible
massive representation of the Poincare group if the following
conditions are satisfied
\begin{eqnarray}
&&
(i\gamma^\nu\partial_\nu-m)\psi_\mu=0,
\qquad
\gamma^\mu\psi_\mu=0,
\qquad
\partial^\mu\psi_\mu=0,
\label{irrep-f}
\end{eqnarray}
with $\{\gamma^\mu,\gamma^\nu\}=-2g^{\mu\nu}$.
When we put these equations on an arbitrary curved spacetime
we see that if we do not include the terms with the inverse powers of the
mass then
there is no freedom to add any terms with the curvature and it
unambiguously follows that in curved space equations
(\ref{irrep-f}) take the form
\begin{eqnarray}
(i\gamma^\nu\nabla_\nu-m)\psi_\mu=0,
\qquad
\gamma^\mu\psi_\mu=0,
\qquad
\nabla^\mu\psi_\mu=0.
\label{irrep-fc}
\end{eqnarray}

Let us find what spaces do not give supplementary equations in addition to
(\ref{irrep-fc}).
For this purpose we take
the divergence of the mass-shell equation
and suppose that equations on $\psi_\mu$ (\ref{irrep-fc}) are satisfied\footnote{Our definition of the curvature tensor is $R^{\alpha}_{\;\;\beta\mu\nu}=\partial_{\mu}\Gamma^{\alpha}_{\;\;\beta\nu}-
\partial_{\nu}\Gamma^{\alpha}_{\;\;\beta\mu}
+\Gamma^{\lambda}_{\beta\nu}\Gamma^{\alpha}_{\;\;\lambda\mu}
-\Gamma^{\lambda}_{\;\;\beta\mu}\Gamma^{\alpha}_{\;\;\lambda\nu}$}
\begin{eqnarray}
0&=&\nabla^\mu(i\gamma^\nu\nabla_\nu-m)\psi_\mu
=i\gamma^\nu[\nabla^\mu,\nabla_\nu]\psi_\mu
=i\tilde{R}^{\mu\nu}\gamma_\nu\psi_{\mu}
-\frac{i}{4}R^{\alpha\beta\mu\nu}\gamma_\nu\gamma_{\alpha\beta}\,\psi_\mu
,
\label{compat}
\end{eqnarray}
where
$\gamma_{\alpha\beta}=\gamma_{[\alpha\beta]}=1/2(\gamma_\alpha\gamma_\beta-\gamma_\beta\gamma_\alpha)$
and
$\tilde{R}_{\mu\nu}=R_{\mu\nu}-\frac{1}{d}g_{\mu\nu}R$.
Extracting from $\gamma_\nu\gamma_{\alpha\beta}$ totally antisymmetric part
$\gamma_{\nu\alpha\beta}=1/6(\gamma_\nu\gamma_\alpha\gamma_\beta\pm\ldots)$
\begin{eqnarray}
\gamma_\nu\gamma_{\alpha\beta}
=
\gamma_{\nu\alpha\beta}+\gamma_\alpha g_{\beta\nu}-\gamma_\beta g_{\alpha\nu}
\end{eqnarray}
and substituting into (\ref{compat}) one gets
\begin{eqnarray}
\frac{i}{2}\tilde{R}^{\mu\nu}\gamma_\nu\psi_{\mu}=0.
\label{compat-2}
\end{eqnarray}
From (\ref{compat-2}) we see that if the traceless part of the Ricci
tensor is zero $R_{\mu\nu}-\frac{1}{d}g_{\mu\nu}R=0$ then equations
(\ref{irrep-fc}) are compatible with each other. Next from
$R_{\mu\nu}=\frac{1}{d}g_{\mu\nu}R$ follows that $R=const$, while
the Weyl tensor $C_{\mu\alpha\beta\nu}$ remains arbitrary. That is
equations (\ref{irrep-fc}) are compatible with each other in the
Einstein space. In the further part of the paper we construct
Lagrangian for spin-$\frac{3}{2}$ field in such space using the new
aspect of BRST approach elaborated in \cite{09120611}.


First of all we note that in the Einstein spaces the equations of
motion (\ref{irrep-fc}) can be modified in such way that a parameter
with dimension of length appears in the equations. In this case it
is reasonable to construct the Lagrangian leading to mass-shell
equations, formally coinciding with equations which define the
irreducible representations of the AdS group \cite{9802097}. Thus
the mass-shell equations  which are expected to be resulted from the
Lagrangian (up to gauge fixing) should have the form
\begin{eqnarray}
[\;i\gamma^\sigma\nabla_\sigma-m-\tfrac{d-2}{2}r^\frac{1}{2}\;]\psi_\mu=0,
\qquad
\gamma^\mu\psi_\mu=0,
\qquad
\nabla^\mu\psi_\mu=0,
\label{TheEM}
\end{eqnarray}
where $r$ is defined from
$R_{\mu\nu\alpha\beta}=r(g_{\mu\beta}g_{\nu\alpha}-g_{\mu\alpha}g_{\nu\beta})+C_{\mu\nu\alpha\beta}$,
i.e. $R=-rd(d-1)$.

Now we introduce auxiliary Fock space generated by
fermionic\footnote{Of course, in the case under consideration, we
could also use bosonic creation and annihilation operators, but use
of fermionic ones is simpler, cf. \cite{09021471} and
\cite{0703049}.} creation and annihilation operators $a_a^+$, $a_a$
satisfying the anticommutation relations
\begin{eqnarray}
\{a^+_a,a_b\}=\eta_{ab},
\qquad
\eta_{ab}=diag(-,+,+,\cdots,+)
.
\end{eqnarray}
Our further consideration is very close to \cite{09021471},
therefore we will omit some details of the calculations. As usual
the tangent space indices and the curved space indices are converted
one into another with the help of vielbein $e^a_\mu$ which is
assumed to satisfy the relation $\nabla_\nu{}e^a_\mu=0$. Then in
addition to the conventional gamma-matrices
\begin{eqnarray}
\{\gamma_a, \gamma_b\}=-2\eta_{ab},
\end{eqnarray}
we introduce a set of $d+1$ Grassmann
odd objects \cite{0603212,0703049,09021471} which obey the following gamma-matrix-like conditions
\begin{eqnarray}
\{\tilde{\gamma}{}^a,\tilde{\gamma}{}^b\}=-2\eta^{ab},
&\qquad&
\{\tilde{\gamma}{}^a,\tilde{\gamma}\}=0,
\qquad
\tilde{\gamma}{}^2=-1
\label{tildedgamma}
\end{eqnarray}
and connected with
the ``true'' (Grassmann even)
gamma-matrices by the relation
\begin{eqnarray}
\gamma^a&=&\tilde{\gamma}{}^a\tilde{\gamma}
           =-\tilde{\gamma}\tilde{\gamma}{}^a.
\label{truegamma}
\end{eqnarray}
After this we define derivative operator
\begin{eqnarray}
D_\mu=\partial_\mu+\omega_\mu{}^{ab}M_{ab},
\qquad
M_{ab}=
{\textstyle\frac{1}{2}}(a_a^+a_b-a_b^+a_a)
-{\textstyle\frac{1}{8}}(
\tilde{\gamma}_a\tilde{\gamma}_b-\tilde{\gamma}_b\tilde{\gamma}_a),
\end{eqnarray}
which acts on an arbitrary state vector in the Fock
space
\begin{eqnarray}
\label{PhysState}
|\psi\rangle=\sum_{n=0}|\psi_n\rangle,
&\qquad&
|\psi_n\rangle=
a^{+\mu_1}\cdots a^{+\mu_n}\psi_{\mu_1\cdots\mu_n}(x)|0\rangle,
\end{eqnarray}
as the covariant derivative\footnote{We assume that
$\partial_\mu{}a_a^+=\partial_\mu{}a_a=\partial_\mu|0\rangle=0$.}
\begin{eqnarray}
D_\mu|\psi_n\rangle
&=&
a^{\mu_1+}\ldots a^{\mu_n}
(\nabla_\mu\psi_{\mu_1\ldots\mu_n})|0\rangle.
\end{eqnarray}
As a result equations (\ref{TheEM}) can be realized in the operator form
\begin{eqnarray}
\label{TheEM-op}
t_0|\psi_1\rangle=0,
\qquad
t_1|\psi_1\rangle=0,
\qquad
l_1|\psi_1\rangle=0,
\end{eqnarray}
where
\begin{eqnarray}
t_0=i\tilde{\gamma}^\mu D_\mu
+\tilde{\gamma}(r^\frac{1}{2}g_0-m),
\qquad
g_0=a_\mu^+a^\mu-{\textstyle\frac{d}{2}},
\qquad
t_1=\tilde{\gamma}^\mu a_\mu,
\qquad
l_1=-ia^\mu D_\mu.
\end{eqnarray}

Lagrangian construction within the BRST approach
\cite{brst-1,brst-2,0603212,0703049,09021471,09120611}
demands that we must have at hand a set of operators which is
invariant under Hermitian conjugation and which forms an algebra
\cite{brst-1,brst-2,0603212,0703049,09021471,09120611}.
In order to determine the Hermitian
conjugation properties of the constraints we define the following
scalar product
\begin{equation}
\label{sproduct}
\langle\tilde{\Psi}|\Phi\rangle
=
\int d^dx \sqrt{-g} \sum_{n,\,k=0}
\langle0|
\Psi^+_{\nu_1\ldots\,\nu_k}(x) \tilde{\gamma}^0
a^{\nu_k}\ldots\,a^{\nu_{1}}
a^{+\mu_1}\ldots\,a^{+\mu_n}
\Phi_{\mu_1\ldots\,\mu_n}(x)
|0\rangle
.
\end{equation}
As a result we see that constraint $t_0$ is Hermitian and
the two other are non-Hermitian\footnote{We assume that
$(\tilde{\gamma}^\mu)^+=\tilde{\gamma}^0\tilde{\gamma}^\mu\tilde{\gamma}^0$,
$(\tilde{\gamma})^+=\tilde{\gamma}^0\tilde{\gamma}\tilde{\gamma}^0=-\tilde{\gamma}$.}
\begin{eqnarray}
t_1^+=a^+_\mu\tilde{\gamma}^\mu,
&\qquad&
l_1^+=-ia^{\mu+}D_\mu.
\end{eqnarray}
Thus set of operators $t_0$, $t_1$, $t_1^+$, $l_1$, $l_1^+$ is invariant under Hermitian conjugation.
Then for constructing the BRST operators the underlying set of
operators must form an algebra. Note that the nilpotency condition
of the BRST operators is needed for existing of gauge symmetry. As it
is known if we consider half-odd spin-s field and decompose the gauge parameter
$|\Lambda_n\rangle$ ($s=n+1/2$) in series of creation operators,
then maximal tensorial rank of gauge parameters
$|\lambda_{k}\rangle=a^{+\mu_1}\cdots{}a^{+\mu_{k}}\lambda_{\mu_1\cdots\mu_{k}}|0\rangle$,
entering in $|\Lambda_n\rangle$ is $k=n-1$ (see e.g. \cite{0603212,0703049,09021471}). Therefore
if we want to construct Lagrangian for a particular half-odd spin-s field it
is enough that this set of operators forms algebra only on states
$|\lambda_{k}\rangle$
with $k<s-1/2$.

Now in order to an algebra we add to the set of operators all the
operators generated by the (anti)commutators of $t_0$, $t_1$, $l_1$,
$t_1^+$, $l_1^+$, but unlike the case of arbitrary spin
\cite{0603212,0703049,09021471} the algebra must be closed on states
$|\lambda_0\rangle=\lambda(x)|0\rangle$ only. Doing similar
considerations as in \cite{09120611} we arrive to the conclusion
that there should be added the following three operators
\begin{eqnarray}
&&
l_0=D^2-m^2+r\bigl(-g_0^2+g_0+t_1^+t_1+{\textstyle\frac{d(d+1)}{4}}\bigr),
\\
&&
g_0=a_\mu^+a^\mu-{\textstyle\frac{d}{2}},
\qquad
g_m=m,
\end{eqnarray}
where
$D^2=g^{\mu\nu}(D_{\mu}D_\nu-\Gamma_{\mu\nu}^{\sigma}D_\sigma)$.
As a result set of operators
$o_i=(t_0, l_0, t_1, l_1, t_1^+, l_1^+, g_0, g_m)$
is invariant under Hermitian conjugation and forms an algebra on states
$|\lambda_0\rangle$ in the Einstein space.
Note that the form of the algebra coincides with those obtined in
\cite{09021471}, therefore we borrow from there all the results needed for
construction of Lagrangian for spin-$\frac{3}{2}$ field in the space under
consideration.

First, since the algebra contains operators $g_0$ and $g_m$ which are not
constraints neither in the bra- nor in the ket-vector space then we should
construct new expressions $o_i\to{}O_i$ for the operators of the
algebra,
so that the operators which
are not constraints don't give supplementary equations on the physical field
and form an algebra.
We borrow the result for these new expressions for the operators from \cite{09021471}
\begin{eqnarray}
&&
T_0=i\tilde{\gamma}^\mu D_\mu-\tilde{\gamma}m_0-2m_1f^+b
+\frac{r}{2m_1}\Bigl(b^+b+2h\Bigr)b^+f,
\label{en-1}
\\
&&
T_1^+=a^+_\mu\tilde{\gamma}^\mu+b^+,
\qquad
L_1^+=-ia^{+\mu}D_\mu+m_1f^+-\frac{r}{4m_1}\,b^{+2}f,
\qquad
G_m=0,
\\
&&
L_0=D^2-m_0^2+r(g_0+t_1^+t_1+{\textstyle\frac{d(d+1)}{4}})
-r(b^+b+2h)b^+b-2r(b^+b+h+{\textstyle\frac{1}{2}})f^+f,
\\
&&
T_1=\tilde{\gamma}^\mu a_\mu+\tilde{\gamma}\frac{m_0}{m_1}f+(2f^+f+b^+b+2h)b,
\qquad
G_0=g_0+b^+b+f^+f+h,
\\
&&
L_1=-ia^\mu D_\mu+\tilde{\gamma}m_0b+m_1f^+b^2+\frac{m_0^2}{m_1}f
-\frac{r}{m_1}(h+{\textstyle\frac{1}{2}})(b^+b+h)f
-\frac{r}{4m_1}b^{+2}b^2f
,
\label{en-5}
\end{eqnarray}
where in case of spin-$\frac{3}{2}$ field one should take $h=d/2-1$ and
$m_0=m+r^{\frac{1}{2}}(d-2)/2$.
In eqs. (\ref{en-1})--(\ref{en-5})
we have introduced one pair of fermionic $f^+$, $f$ and
one pair of bosonic $b^+$, $b$ creation and annihilation
operators with the standard (anti)commutation relations
\begin{eqnarray}
\{f^+,f\}=1,&\qquad& [b^+,b]=1.
\label{bf-add}
\end{eqnarray}
Also expressions (\ref{en-1})--(\ref{en-5})
contain arbitrary (nonzero) constant $m_1$ with dimension of mass. Its value remains
arbitrary and it can be expressed from the other parameters of
the theory $m_1=f(m,r)\neq0$. The arbitrariness of this
parameter does not influence on the reproducing of the equations
of motion for the physical field (\ref{TheEM}).

Note that the new expressions for the operators do not obey the usual properties
\begin{align}
&
(T_0)^+\neq T_0
&&
(L_0)^+\neq L_0,
&&
(T_1)^+\neq T_1^+,
&&
(L_1)^+\neq L_1^+
\end{align}
if one uses the standard rules of Hermitian conjugation for the new
creation and annihilation operators
\begin{equation}
(b)^+=b^+,
\qquad
(f)^+=f^+.
\end{equation}
To restore the proper Hermitian conjugation properties for the
additional parts we change the scalar product in the Fock space generated by the
new creation and annihilation
operators  as follows:
\begin{eqnarray}
\langle\tilde{\Psi}_1|\Psi_2\rangle_{\mathrm{new}} =
\langle\tilde{\Psi}_1|K|\Psi_2\rangle\,, \label{newsprod}
\end{eqnarray}
for any vectors $|\Psi_1\rangle, |\Psi_2\rangle$ with some operator
$K$. Since the problem with the proper Hermitian conjugation of the
operators is only in $(b^+, f^+)$-sector of the Fock space, the
modification of the scalar product takes place only in this sector.
Therefore operator $K$ acts as a unit operator in the entire Fock
space, but for the $(b^+, f^+)$-sector where the operator has the
form
\begin{eqnarray}
K&=&\sum_{k=0}^{\infty}\frac{C_h(k)}{k!}\biggl[\;
|0,k\rangle\frac{1}{2h+k}\langle0,k|
\;+\;|1,k\rangle\frac{2m_0^2-rh(2h{+}k{+}1)}{4h\,m_1^2}\langle1,k|
\nonumber
\\
&&\hspace*{7em}{}
+|1,k\rangle\frac{\tilde{\gamma}m_0}{2h\,m_1}\langle0,k{+}1|
\;+\;|0,k{+}1\rangle\frac{\tilde{\gamma}m_0}{2h\,m_1}\langle1,k|
\;\;\biggr]
,
\label{K}
\end{eqnarray}
where
\begin{eqnarray}
&&
C_h(k)=2h\,(2h+1)\cdot\ldots\cdot(2h+k-2)(2h+k-1)(2h+k),
\\
&&
|0,k\rangle=(b^+)^k|0\rangle,
\qquad
|1,k\rangle=f^+(b^+)^k|0\rangle.
\end{eqnarray}

Next step is constructing the BRST operator on the base of algebra generated
by (\ref{en-1})--(\ref{en-5}). Its explicit form can be found in
\cite{09021471}.
Then using the found BRST operator one constructs (up to an overall factor)
Lagrangian and gauge transformation for spin-$\frac{3}{2}$ field in the Einstein space
(the details can be found in e.g. \cite{09021471,0703049})
\begin{eqnarray}
{\cal{}L}
&=&
\langle\tilde{\chi}^{0}|K\tilde{T}_0|\chi^{0}\rangle
+
\frac{1}{2}\,\langle\tilde{\chi}^{1}|K\bigl\{
   \tilde{T}_0,q_1^+q_1\bigr\}|\chi^{1}\rangle
+
\langle\tilde{\chi}^{0}|K\Delta{}Q|\chi^{1}\rangle
+
\langle\tilde{\chi}^{1}|K\Delta{}Q|\chi^{0}\rangle
,
\label{L1}
\end{eqnarray}
\begin{eqnarray}
\delta|\chi^{0}\rangle=\Delta{}Q|\Lambda\rangle
&\qquad&
\delta|\chi^{1}\rangle=\tilde{T}_0|\Lambda\rangle,
\label{GT}
\end{eqnarray}
where
\begin{eqnarray}
\tilde{T}_0&=&
T_0
+2i(\eta_1^+p_1-\eta_1p_1^+)
+\frac{r}{2}(2g_0-G_0)(q_1\mathcal{P}_1^++q_1^+\mathcal{P}_1)
\\
\Delta{}Q&=&
\eta_1^+T_1+\eta_1T_1^+
+q_1^+L_1+q_1L_1^+
-\frac{r}{4}\Bigl[
q_1(2t_1^+-T_1^+)+q_1^+(2t_1-T_1)
\Bigr]
(q_1\mathcal{P}_1^++q_1^+\mathcal{P}_1)
\end{eqnarray}
and
\begin{align}\label{chi32}
&
|\chi^0\rangle
=
\left[
-ia^{+\mu}\psi_{\mu}(x)+f^+\varphi(x)+b^+\tilde{\gamma}\psi(x)
\right]|0\rangle,
&&
|\chi^1\rangle
=
\left[\mathcal{P}^{+}_1\chi_1(x)-ip_1^+\tilde{\gamma}\chi(x)
\right]|0\rangle,
\\
&
|\Lambda\rangle
=
\left[
\mathcal{P}^{+}_1\tilde{\gamma}\lambda_1(x)-ip_1^+\lambda(x)
\right]|0\rangle
.
\label{l001}
\end{align}
Here $q_1$, $q_1^+$ are bosonic and $\eta_1^+$, $\eta_1$ are fermionic
ghost ``coordinates'' corresponding to their
canonically conjugate ghost ``momenta''
$p_1^+$, $p_1$, ${\cal{}P}_1$, ${\cal{}P}_1^+$
obeying the (anti)commutation relations
\begin{eqnarray}
\{\eta_1,\mathcal{P}_1^+\}=
\{\eta_1^+,\mathcal{P}_1\}=1,
&\quad&
[q_1,p_1^+]=[q_1^+,p_1]=i
\label{ghosts}
\end{eqnarray}
and possess the standard  ghost number distribution
$gh(\mathcal{C}^i)=-gh(\mathcal{P}_i)=1$
and act on the vacuum state as follows
\begin{equation}
(q_1, p_1, \eta_1, \mathcal{P}_1 )
|0\rangle
=0.
\label{ghostvac}
\end{equation}

Substituting (\ref{chi32}), into (\ref{L1}), we
find the action   (up to an overall factor) for a spin-$\frac{3}{2}$ field
in the Einstein spacetime in the component form
\begin{eqnarray}
S
&=&
\int d^dx \sqrt{-g}
\left[\bar{\psi}^\mu\Bigl\{
\bigl[i\gamma^\nu\nabla_\nu-m_0\bigr]\psi_\mu -\nabla_\mu\chi
-i\gamma_\mu\chi_1 \Bigr\} \right.\nonumber
\\
&&{}
-
\Bigl[
(d-2)\bar{\psi}
+
\frac{m_0}{m_1}\bar{\varphi}
\Bigr]
\Bigl\{
\bigl[i\gamma^\mu\nabla_\mu+m_0\bigr]\psi
+\frac{r(d-1)}{2m_1}\;\varphi
+\chi_1
\Bigr\}
\nonumber
\\
&&{}
+
\Bigl[
\frac{M^2}{m_1^2}\bar{\varphi}
-\frac{m_0}{m_1}\bar{\psi}
\Bigr]
\Bigl\{
\bigl[i\gamma^\sigma\nabla_\sigma-m_0\bigr]\varphi
+2m_1\psi
-m_1\chi
\Bigr\}
\nonumber
\\
&&{}
-
\bar{\chi}
\Bigl\{
\bigl[
i\gamma^\mu\nabla_\mu
+m_0
\bigr]
\chi
+\chi_1
-\nabla^\mu\psi_\mu
-m_0\psi
+\frac{M^2}{m_1}\;\varphi\Bigr\}
\nonumber
\\
&&{}
\left.
+\bar{\chi}_1
\Bigl\{ i\gamma^\mu\psi_\mu -(d-2)\psi
-\chi -\frac{m_0}{m_1}\varphi \Bigr\}\right] \label{L3/2} ,
\end{eqnarray}
where $M^2=m_0^2-\frac{1}{4}\;r(d-1)(d-2)$ and $m_0=m+r^{\frac{1}{2}}(d-2)/2$.
Substituting (\ref{chi32}), (\ref{l001})
into (\ref{GT}), we find the gauge transformations in the component form
\begin{align}
&
\delta\psi_\mu
=\nabla_\mu\lambda+i\gamma_\mu\lambda_1,
&&
\delta\psi=\lambda_1,
\hspace*{7em}
\delta\varphi=m_1\lambda
\\
&
\delta\chi=
\Bigl[
i\gamma^\mu\nabla_\mu
-m_0
\Bigr]\lambda
+2\lambda_1,
&&
\delta\chi_1
=
-\Bigl[
i\gamma^\mu\nabla_\mu
+m_0
\Bigr]
\lambda_1
-\frac{r}{2}(d-1)
\lambda
.
\label{gtr}
\end{align}

Beginning with the Lagrangian (\ref{L3/2}) we can obtain the other
Lagrangians for spin-$\frac{3}{2}$ field containing less number of
involved fields. Let us present the action in terms of one basic
field $\psi_\mu$. To this end, we get rid of the fields $\varphi$,
$\psi$, by using their gauge transformations and the gauge
parameters $\lambda$, $\lambda_1$, respectively. Having expressed
the field $\chi$, using the equation of motion
$\chi=i\gamma^\mu\psi_\mu$, we see that the terms with the
Lagrangian multiplier $\chi_1$ disappear. As a result, we obtain
\begin{eqnarray}
\mathcal{L} &=&
\bar{\psi}^\mu(i\gamma^\sigma\nabla_\sigma-m_0)\psi_\mu
-i\bar{\psi}^\mu(\gamma_\nu\nabla_\mu+\gamma_\mu\nabla_\nu)\psi^\nu
-\bar{\psi}^\nu\gamma_\nu(i\gamma^\sigma\nabla_\sigma+m_0)\gamma^\mu\psi_\mu
. \label{RSL}
\end{eqnarray}
Thus we have obtained the Lagrangian for massive spin-$\frac{3}{2}$ field
in $d$-dimensional Einstein space only in term of the basic field. In
the massless case $m=0$ ($m_0=r^{\frac{1}{2}}(d-2)/2$) and
Lagrangian (\ref{RSL}) becomes invariant under gauge transformation
\begin{eqnarray}
 \delta\psi_\mu=\nabla_\mu\lambda-\frac{ir^{\frac{1}{2}}}{2}\gamma_\mu\lambda
.
\end{eqnarray}


Our next aim is discussing the problem of causality for  massive
spin-$\frac{3}{2}$ field with Lagrangian (\ref{RSL}) in Einstein
space.
Consideration is based on the Velo and Zwanziger method \cite{Velo}
reformulated for the theories in curved spacetime.

We begin with a brief outline of the method.  If one has a system of the
first order differential equations for a set of fields $\varphi^B$
\begin{eqnarray}
G^A_{B}{}^\mu\partial_\mu\varphi^B+\ldots=0,
\qquad
\mu,\nu=0,\ldots,{}d-1
\label{System}
\end{eqnarray}
then to verify that the system (\ref{System}) describes hyperbolic propagation
one should check that all solutions $n_0(n_i)$, $(i=1,\ldots{}d-1)$  of the algebraic equation
\begin{eqnarray}
\det(G^A_B{}^{\mu}n_\mu)=0
\label{Det}
\end{eqnarray}
are real for any given real set of $n_i$.
The hyperbolic system is called causal if there are no timelike vectors among
solutions $n_\mu$ of (\ref{Det}).

In many physical cases (including spin-$\frac{3}{2}$ field) equation (\ref{Det})
fulfills identically. In this case one should replace the initial
system of equations by another equivalent system of equations
supplemented by constraints at a given initial time. Then the above
analysis should be applied to this new system.

Let us turn to our spin-$\frac{3}{2}$ field described by Lagrangian (\ref{RSL}).
The equations of motion are
\begin{eqnarray}
E_{\mu}&\equiv&
(i\gamma^\sigma\nabla_\sigma-m_0)\psi_\mu
-i(\gamma_\nu\nabla_\mu+\gamma_\mu\nabla_\nu)\psi^\nu
-\gamma_\mu(i\gamma^\sigma\nabla_\sigma+m_0)\gamma^\nu\psi_\nu
=0.
\label{EM}
\end{eqnarray}
If we consider equation (\ref{Det}) for equations (\ref{EM}) then we
find that it fulfills identically. Therefore one should replace
equations (\ref{EM}) by another equivalent system of equations with
constraints on initial data. It can be done by the same method as in
\cite{Velo} and we will not repeat all the steps and proofs. The
system of equations equivalent to (\ref{EM}) is
\begin{eqnarray}
E_\mu+i\gamma_\mu C +i\nabla_\mu D +m_0\gamma_\mu D
=(i\gamma^\sigma\nabla_\sigma-m_0)\psi_\mu=0,
\label{EM-2}
\end{eqnarray}
where
\begin{eqnarray}
C&=&
\frac{ir(d-2)\gamma^\mu E_\mu-4m_0\nabla^\mu E_\mu}{4m_0^2-r(d-2)^2}
=\nabla^\mu\psi_\mu+\gamma^\sigma\nabla_\sigma\gamma^\nu\psi_\nu
,
\\
D&=&\frac{4}{d-1}\;\frac{m_0\gamma^\mu E_\mu+i(d-2)\nabla^\mu E_\mu}{4m_0^2-r(d-2)^2}
=\gamma^\nu\psi_\nu
\end{eqnarray}
and it is supplemented by constraints\footnote{In curved spacetime the question of causality should be considered locally at some arbitrary point $x_0$ choosing around $x_0$ the Riemann normal coordinates.
In this case $E_0$ appears as a constraint.} at an initial time (say $t=0$)
\begin{eqnarray}
E_{0}|_{t=0}=0,
\qquad
\gamma^\mu\psi_\mu|_{t=0}=0.
\end{eqnarray}
Thus, like in the flat case, the equations for spin-$\frac{3}{2}$
field (\ref{EM}), following from Lagrangian (\ref{RSL}), are
hyperbolic and causal for the background under consideration.


To conclude, we have shown that the BRST approach to higher spin
field theories yields consistent Lagrangian construction for
spin-$\frac{3}{2}$ field in general Einstein space.  As usual in the
BRST approach, massive spin-$\frac{3}{2}$ Lagrangian is obtained in
gauge invariant form with suitable St\"uckelberg auxiliary fields.
Analysis has been based on some hidden aspect of the BRST-BFV
construction which was explored in context of higher spin filed
theory in \cite{09120611}. We saw that the BRST approach perfectly
works if one requires that the BRST operator $Q_{s}$ is nilpotent in
weak sense, i.e. $Q_{s}^2|\Lambda_s\rangle=0$ for some spin $s$,
where $|\Lambda_s\rangle$ is a Fock space valued gauge parameter.
One can show that above condition is realized only for exceptional
spins $s=1,\frac{3}{2},2$. The cases $s=1,2$ have been considered in
\cite{09120611}. Here we have studied the last exceptional case
$s=\frac{3}{2}$. Also we have proved that in Einstein space the
spin-$\frac{3}{2}$ field propagation is hyperbolic and causal. As we
know such proof was known before only for the constant curvature
space.

{\bf {Acknowledgements}}. The authors are grateful to P.M. Lavrov for
discussions on BRST-BFV construction. The work is partially
supported by the RFBR grant, project No.\ 09-02-00078, the
RFBR-Ukraine grant, project No.\ 10-02-90446, grant for LRSS,
project No.\ 3558.2010.2. The work of I.L.B. is also partially
supported by the RFBR-DFG grant, project No.\ 09-02-91349 and the
DFG grant, project No. 436 RUS 113/669/0-4.

\end{document}